\documentclass{new_tlp}

\usepackage{mathptmx}
\usepackage{latexsym}
\usepackage{color}
\usepackage{ifthen}
\usepackage{multirow}
\usepackage[utf8]{inputenc} % Caracteres españoles
\usepackage{graphicx} % Inclusión de imágenes
\usepackage{url} % Direcciones web, correos, ...
\usepackage[normalem]{ulem} % Tachado de texto
\usepackage{calc}
\usepackage{latexsym}
\usepackage{stmaryrd} % For \llbracket and \rrbracket

\newcommand{\myurl}[1]{{\fontsize{9}{9}\url{#1}}}
\newcommand{\mysf}[1]{{\fontsize{10}{10}\textsf{#1}}}
\newcommand{\PL}{{\mysf{Prolog}}}

\newcommand{\BPL}{{\mysf{Bousi$\sim$Prolog}}}

\newcommand{\fDES}{{\mysf{FuzzyDES}}}

\newcommand{\SWIPL}{{\mysf{SWI-Prolog}}}

  {\vspace{0.5\baselineskip}\small\begin{list}{\labelwidth=0pt\itemindent=0pt}\item}%
  {\end{list}\vspace{0.5\baselineskip}}

\newfont{\msbm}{msbm10 scaled 1000}

\newcommand{\ol}[1]{\overline{#1}}  % sequence of objects
\def \tuple#1{\langle #1 \rangle} %%% tuple
\newcommand{\T}{\mbox{\footnotesize $\triangle$}}
\newcommand{\wmgu}{{\sf\small wmgu}_{\cR}^{\lambda}} %% Example: $\wmgu(s,t)$
\newcommand{\revto}{\leftarrow}

\newcommand{\sld}[1]{\stackrel{\!\!\! #1 \;}{{\Rightarrow}_{\mbox{\tiny WSLD}}}}

\def\defemb#1#2{\expandafter\def\csname #1\endcsname
                              {\relax\ifmmode #2\else\hbox{$#2$}\fi}}
\defemb{cA}{{\cal A}}
\defemb{cB}{{\cal B}}
\defemb{cC}{{\cal C}}
\defemb{cD}{{\cal D}}
\defemb{cE}{{\cal E}}
\defemb{cF}{{\cal F}}
\defemb{cG}{{\cal G}}
\defemb{cH}{{\cal H}}
\defemb{cI}{{\cal I}}
\defemb{cJ}{{\cal J}}
\defemb{cK}{{\cal K}}
\defemb{cL}{{\cal L}}
\defemb{cM}{{\cal M}}
\defemb{cN}{{\cal N}}
\defemb{cO}{{\cal O}}
\defemb{cP}{{\cal P}}
\defemb{cQ}{{\cal Q}}
\defemb{cR}{{\cal R}}
\defemb{cS}{{\cal S}}
\defemb{cT}{{\cal T}}
\defemb{cU}{{\cal U}}
\defemb{cV}{{\cal V}}
\defemb{cX}{{\cal X}}
\defemb{cY}{{\cal Y}}
\defemb{cZ}{{\cal Z}}

\defemb{wR}{\mbox{\footnotesize $\widehat{\cal R}$}}
\long\def\comment#1{}

\newcommand{\todoFlag}{ON} 
\newcommand{\todo}[1]{\ifthenelse{\equal{\todoFlag}{ON}}{\fcolorbox{red}{yellow}{
			\begin{minipage}{0.9\linewidth}#1\end{minipage}
	}}{}}

\newcommand{\tachar}[1]{\ifthenelse{\equal{\todoFlag}{ON}}{{\color{red}\sout{#1}}}{}}
\newcommand{\myfontcodesize}{\fontsize{10}{10}}
\newcommand{\mytt}[1]{{\myfontcodesize \texttt{\textbf{#1}}}}
\title[Implementing WordNet Lexical Similarity Measures] 
        {Implementing WordNet Measures of Lexical Semantic Similarity in a Fuzzy Logic Programming System\thanks{Work partially funded by the State Research Agency (AEI) of the Spanish Ministry of Science and Innovation under grant PID2019-104735RB-C42 (SAFER), by the Spanish Ministry of Economy and Competitiveness, under the grants TIN2016-76843-C4-2-R (MERINET), TIN2017-86217-R (CAVI-ART-2), and by the Comunidad de Madrid, under the grant S2018/TCS-4339 (BLOQUES-CM), co-funded by EIE Funds of the European Union.}
        	}

\author[Pascual Juli\'{a}n-Iranzo and Fernando S\'{a}enz-P\'{e}rez]
	{{PASCUAL JULI\'AN-IRANZO}\\
	Dept. of Information Technologies and Systems, University of Castilla-La Mancha, \\
	13071 Ciudad Real, Spain\\
	\email{Pascual.Julian@uclm.es}
	\and
	{FERNANDO S\'AENZ-P\'EREZ}\\
	Faculty of Computer Science, Complutense University of Madrid, \\
	28040 Madrid, Spain \\
	\email{fernan@sip.ucm.es}
	}

\pagerange{\pageref{firstpage}--\pageref{lastpage}}
\begin{document}
	%\nocite{*}% includes all entries of BibTeX database into the list of references.

	\label{firstpage}
	
	\maketitle

\begin{abstract}
    This paper introduces techniques to  integrate WordNet into a Fuzzy Logic Programming system. Since WordNet relates words but does not give graded information on the relation between them, we have implemented standard similarity measures and new directives allowing the proximity equations linking two words to be generated with an approximation degree. Proximity equations are the key syntactic structures which, in addition to a weak unification algorithm, make a flexible query-answering process possible in this kind of programming language.
This addition widens the scope of Fuzzy Logic Programming, allowing certain forms of lexical reasoning, and reinforcing Natural Language Processing applications.
\begin{center}
\emph{Under consideration in Theory and Practice of Logic Programming (TPLP)}
\end{center}
\end{abstract}


%%%%%%%%%%%%
%\tableofcontents

%%%%%%%%%%%%%%%%%%%%%%%%%%%%%%%%%%%%%%%%%%%%%%%%%%%%%%%%%%%%%%%

%%%%%%%%%%%%%%%%%%%%%%%%%%%%%%%%%%%%%%%
\section{Introduction and Motivation} \label{sec-intro}

{\em Fuzzy Logic Programming} \cite{Lee72} integrates concepts coming from fuzzy logic \cite{Zad65} into logic programming \cite{vEK76} in order to deal with the essential vagueness of some problems by using declarative techniques.  
In recent years there has been renewed interest in this field, involving multiple lines of work. When the fuzzy unification algorithm is weakened using a similarity relation (i.e., a reflexive, symmetric, transitive, fuzzy binary relation) the approach is usually called {\em Similarity-based Logic Programming} \cite{FF99,FF02,LSS01,Ses02}. 

We have extended Similarity-based Logic Programming by introducing new theoretical concepts and developing two fuzzy logic programming systems: \BPL\ (\mysf{BPL} for short) \cite{RJ14JIFS,JR17FSS} and \fDES\ \cite{JS17,JS18a}.
Their syntax is based on the clausal form, and they embody a {\em Weak SLD (WSLD) resolution} operational semantics, which uses a fuzzy unification algorithm based on the concept of {\em proximity relation}  (i.e., a fuzzy binary relation supporting unification that, although reflexive and symmetric, is not necessarily transitive) \cite{JR15FSS,JS18b}.
A proximity relation is defined by {\em proximity equations}, denoted by $a\sim b = \alpha$, whose
intuitive reading is that two constants (either $n$-ary function symbols or $n$-ary predicate symbols), $a$ and $b$, are approximate or similar with a certain degree $\alpha$.

%%\begin{example}
For instance, assume a deductive database that stores information about people and their family relationships encoded using the \BPL\ language (see Fig.~\ref{fig_BPL_program}).   
\begin{figure}[t]%[h]
\begin{center}
\begin{verbatim}
%% PROXIMITY EQUATIONS          
ancestor~ascendant=1.0.    ancestor~progenitor=0.9.              
    
%% FACTS
father(abraham,isaac).     father(isaac,esau).     father(isaac,jacob).      
mother(sara,isaac).        mother(rebeca,jacob).   mother(rebeca,esau). 
    				
%% RULES
direct_ancestor(X,Y) :- father(X,Y); mother(X,Y).
        		
ancestor(X,Z) :- direct_ancestor(X,Z).
ancestor(X,Z) :- direct_ancestor(X,Y), ancestor(Y,Z).
\end{verbatim}
\caption{A BPL program fragment.}
\label{fig_BPL_program}
\end{center} 
\end{figure}
%%%%
In a \PL\ system (without proximity equations),  asking about the progenitors of {\small\tt isaac} with the query {\small\tt progenitor(X,isaac)} produces no answer. However, \BPL\ answers {\small\tt X=abraham with 0.9} and {\small\tt X=sara with 0.9} thanks to its proximity-based unification algorithm. Since we have specified that {\small\tt progenitor} is close to {\small\tt ancestor} with degree 0.9, these two terms can ``weakly'' unify with approximation degree 0.9, leading
to a refutation.
%%\end{example}

Here, the proximity equations are axiomatically given by the programmer. It would be interesting if the system could provide assistance through its connection to a lexical resource such as WordNet \cite{Fel98,Fel06,Mil95}. 
This study, therefore, deals with the generation of the set of proximity equations both automatically and with a minimal intervention by the  programmer. However, the motivation for integrating WordNet into our logic systems goes beyond this simple help function.
We provide the Prolog implementation, \mytt{wn\_connect}, to connect to WordNet with a number of similarity measures and convenient predicates to be used either in isolation or integrated into \BPL, allowing reasoning with linguistic terms.  
Unlike Distributional Semantic Models such as Word Embeddings or other statistical approaches,  WordNet-based techniques do not require training and facilitate explainability \cite{SWCZ18}.

The usefulness of this proposal lies in its inclusion of applications such as text mining in information retrieval, text classification, and even sentiment analysis~\cite{APASTGK17,BR11,SORH15} (see also \ref{sec-Text-Class}).

%%%%%%%%%%%%%%%%%%%%%%%%%%

%%%%%%%%%%%%%%%%%%%%%%%%%%%%%%%%%%%%%%%%%%%%%%%%%%%%%%%%%%%%%%%
\section{The Lexical Resource WordNet and Prolog} \label{sec-WordNet}

WordNet  is a lexical English language database.  Words of the same syntactic category 
are grouped into sets of synonyms called {\em synsets}. Roughly speaking, the words of a synset have the same meaning in a specific context and they represent a {\em concept} (or word sense). Each synset has a \mytt{synset\_ID}.
Because a word has different senses (meanings), it can belong to different synsets.  WordNet is structured as a semantic net where words are interlinked by lexical relations, and synsets by semantic relations. Synonymy and antonymy are the major lexical relations. Semantic relations serve to build knowledge structures (i.e., networks of synsets --concepts--). 
Nouns, as well as verbs, are interconnected by the {\em hyponymy} relation (IS-A relation), which links specific concepts to more general ones.  
Hypernymy is the opposite relation, that is, a hypernym is a word whose meaning includes a group of other words.
Both  relations are transitive.  Note also that both nouns and verbs are organized as separate hierarchical structures.

WordNet can be accessed either via a web interface or locally. 
There exists a WordNet 3.0 database version released by Eric Kafe which can be found at: \url{https://github.com/ekaf/wordnet-prolog}. The information stored in WordNet is provided as a collection of \PL\ files. 
Each file contains the definition of what is called an {\em operator}, corresponding to a WordNet relation. Files are named \mytt{wn\_<{\em operator}>.pl}, where \mytt{<{\em operator}>} is the name of a specific operation (relation). Therefore, each WordNet relation is represented by a \PL\ predicate which is stored in a separate file and defined by a set of \PL\ facts. The specifications of these predicates are detailed in \cite{Wor06}. We now describe the predicates of greatest interest for this study.

The file \mytt{wn\_s.pl}  contains all the information about words stored in WordNet. It defines the \mytt{s} operator, which has an entry for each word. The structure of the \mytt{s} operator is \mytt{s(Synset\_id, W\_num,Word,Ss\_type,Sense\_number,Tag\_count)}, where \mytt{W\_num}, if given, indicates which word in the synset is being referred to. The words in a synset are numbered serially, starting with 1. The third argument is the word itself (which is represented by a \PL\ atom). The \mytt{Ss\_type} parameter is a one character code indicating the synset type: \mytt{n} (noun); \mytt{v} (verb);  \mytt{a} (adjective); \mytt{s} (satellite adjective) and  \mytt{r} (adverb). The \mytt{Sense\_number} parameter specifies the sense of the word, within the part of speech encoded in the \mytt{Synset\_id}. The higher the sense number, the less common the word. Finally, the \mytt{Tag\_count} indicates the number of times the word sense was found in the sense-tagged text corpus of Semantic Concordances~\cite{MLTB93},  which was generated from the Brown Corpus~\cite{FK79}, using WordNet as a lexicon. The Brown Corpus was inspected word by word, including sense-tags for each one. A higher tag count number means that the word is more common than others with a lower tag count. In Section \ref{sec-WN-access} we illustrate the meaning of some of these parameters through an example.

The file \mytt{wn\_hyp.pl} stores hypernymy  relations in the binary predicate \mytt{hyp(synset\_ID1, synset\_ID2)}   
specifying that the second synset is a hypernym of the first synset. 
This semantic relation only holds for nouns and verbs. 
Because hyponymy is the inverse relation of hypernymy, the operator \mytt{hyp} also specifies that the first synset is a hyponym of the second synset.

%%%%%%%%%%%%%%%%%%%%%%%%%%

%%%%%%%%%%%%%%%%%%%%%%%%%%%%%%%%%%%%%%%%%%%%%%%%%%%%%%%%%%%%%%%%

\section{WordNet and Lexical Semantic Similarity} \label{sec-WordNet-Similarity}

WordNet relates words but does not give their degree of relationship. 
Measuring lexical semantic similarity has many applications for Natural Language Processing (NLP), and its integration into a fuzzy logic programming system such as \BPL\ is appropriate because of its proximity-based operational semantics.
The syntax of our language uses symbols (words) that are endowed with a fuzzy semantics via proximity equations.
We are therefore interested in techniques for measuring the similarity degree between words to facilitate the construction of proximity equations with linguistic criteria.
Semantic similarity quantifies how alike two words are (more precisely: how similar the concepts they denote are). 

Similarity measures  are limited to noun pairs and verb pairs because WordNet organizes nouns and verbs into hyponymy/hypernymy-based hierarchies of concepts (synsets). 

Although a large number of measures of semantic relatedness\footnote{ 
Note that ``lexical semantic relatedness'' is a broad concept that subsumes ``lexical semantic similarity''. There are many different forms in which two words can be related without being similar: for instance, ``car'' and ``petrol'' are closely related, but they are not similar. From a pragmatic point of view, and to distinguish one type of measure from another, it is usual to reserve the name ``relatedness'' for those that measure features other than similarity. 
} 
and similarity have been proposed \cite{BH06}, they are only implemented by a limited number of tools. WordNet::Similarity \cite{PPM04} is perhaps the most prominent. This tool has three similarity measures based on  counting edges  between concepts (PATH, WUP \cite{WP94} and LCH \cite{LC98}), and another three based on information content (RES \cite{Res95}, JCN \cite{JC97} and LIN \cite{Lin98}).

Table~\ref{tab-measures} summarizes some features of these measures, with the measure name in the first column, its type (either counting Edges Based --EB-- or Information Content --IC--) in the second, its description in the third, and its range in the last. In order to understand the description of similarity measures accurately, we introduce the following standard  definitions and notations used when working in the framework of WordNet:

\begin{itemize}
\item 
We differentiate between ``words'' and ``concepts''.  We use the term ``{\em word}'' as shorthand for ``word form,'' and the term ``{\em concept}'' (i.e., ``synset'') to refer to a specific sense or word meaning. Words will be denoted by the letter $w$, and concepts by the letter $c$, possibly with subscripts. 
A concept can also be seen as a word $w$ of type $t$ with sense $s$ and denoted by $w\!:t\!:\!s$ (which we often call  \emph{word term} or \emph{pattern}).

\item Similarity measures use so-called HyperTrees (Hypernym Trees). These are IS-A hierarchies which are a consequence of the hyponymy relation between concepts. 
Despite their name, HyperTrees are not really trees because a concept can be linked to a hypernym concept through different paths. Moreover, in practice, the branches of a HyperTree are manipulated independently as hypernym chains. 
Given a HyperTree, the {\em length of the shortest path} from synset $c_1$ to synset $c_2$ is denoted by $len(c_1,c_2)$. The {\em depth of a node} $c$ is the length of the shortest path from the global root to $c$, i.e., $depth(c) = len(root, c)$.  The ``global root'' is a virtual root that we introduce into the IS-A hierarchy of either nouns or verbs for technical reasons.

\item The {\em least common subsumer} (LCS) of two concepts $c_1$ and $c_2$ is the most specific concept they share as an ancestor. It is denoted by $lcs(c_1, c_2 )$.  An example illustrating the notion of LCS is shown in Section \ref{sec-WN-access}.

\end{itemize}

\begin{table}
\caption{Some similarity measures and their features}
\label{tab-measures}
\begin{minipage}{\textwidth}
\begin{tabular}{cc|c|c|}   
\cline{1-4}
\vspace{-1em}\\
\cline{1-4}
\multicolumn{1}{|@{}c|}{\bf Measure} & {\bf Type} & {\bf Description} & {\bf Range} \\ 
\cline{1-4}
%%%%%%%%%%%%%%%%%%%%%
\multicolumn{1}{|@{}c|}{PATH}
&
EB. 
&
$sim_{PATH}(c_1,c_2) =1/len(c_1, c_2)$
&
$[0,1]$
\\
\cline{1-4}
%%%%%%%%%%%%%%%%%%%%%
\multicolumn{1}{|@{}c|}{WUP}
&
EB. 
&
$sim_{WUP}(c_1,c_2) = \frac{2\times depth(lcs(c_1,c_2))}{Depth(c_1)+Depth(c_2)}$
&
$[0,1]$
\\
\cline{1-4}
%%%%%%%%%%%%%%%%%%%%%
\multicolumn{1}{|@{}c|}{LCH}
&
EB. %%Sim.
&
$sim_{LCH}(c_1,c_2) =   %%$  \hfill $
 -log(\frac{len(c_1,c_2)}{2\times \max\{depth(c)\mid c\in\mbox{\scriptsize WordNet}\}})$
&
$[0,\infty]$
\\
\cline{1-4}
%%%%%%%%%%%%%%%%%%%%%
\multicolumn{1}{|@{}c|}{RES}
&
IC
&
$sim_{RES}(c_1,c_2) = IC(lcs(c_1,c_2))$
&
$[0,\infty]$
\\
\cline{1-4}
%%%%%%%%%%%%%%%%%%%%%
\multicolumn{1}{|@{}c|}{JCN}
&
IC
&
$sim_{JCN}(c_1,c_2)= \frac{1}{IC(c_1) + IC(c_2) - 2\times IC(lcs(c_1,c_2))}$
&
$[0,\infty]$
\\
\cline{1-4}
%%%%%%%%%%%%%%%%%%%%%
\multicolumn{1}{|@{}c|}{LIN}
&
IC
&
$sim_{LIN}(c_1,c_2) = \frac{2\times IC(lcs(c_1,c_2))}{IC(c_1) + IC(c_2)}$
&
$[0,1]$
\\
\cline{1-4}
%\vspace{-1em}\\
\multicolumn{4}{@{}c}{\vspace{-1em}}\\
\cline{1-4}
\end{tabular}
%\vspace{-2\baselineskip}
\end{minipage}
\end{table}%

%%%%%%%%%%%
RES, JCN and  LIN measures are based on the notion of \emph{Information Content} (IC) \cite{Res95}. For a concept $c$,  $IC(c) = -ln(p(c))$, where $p$ is the probability of finding an instance of the concept $c$ in a corpus.
%%%%
In our case, this probability is measured in terms of a relative {\em frequency of use} (or  {\em frequency count}) of the concept $c$ stored in WordNet, which is a measure of the number of times that it occurs in the corpus of Semantic Concordances. Specifically, 
$$
%%%PPP% p(c) = \frac{\mathit{Frequency}(c)}{\mathit{Frequency}(\mathit{Root})}
p(c) = \mathit{Frequency}(c)/\mathit{Frequency}(\mathit{Root})
$$ 
where $\mathit{Frequency}(c)$ is computed by adding the \mytt{Tag\_count} of the concepts subsumed by the concept  $c$ and $\mathit{Root}$ is the concept (virtual or not) on the top of the concept hierarchy.

%%%%%%%%%%%%%%%%%%%%%%%%%%
%%%%%%%%%%%%%%%%%%%%%%%%%%%%%%%%%%%%%%%
\section{Formal Setting} \label{sec-formal}

This section recalls and extends some formal background of \mysf{BPL} and its relation to lexical similarity.
Given a universe $U$, proximity equations extensionally define a \emph{binary fuzzy relation} $\cR: U\times U \longrightarrow [0,1]$. 
A $\lambda$-cut is a user-defined threshold such that $\cR_{\lambda}=\{\tuple{x,y}\mid \cR(x,y)\geq \lambda\}$.  
A fuzzy relation can have some properties attached, for any $e, e_1,e_2,e_3\in U$:
Reflexive ($\cR(e,e)=1$),
Symmetric ($\cR(e_1,e_2)=\cR(e_2,e_1)$)), and
$\T$-Transitive ($\cR(e_1,e_3)\geq \cR(e_1,e_2)\T\cR(e_2,e_3)$),
where the operator $\T$ is an arbitrary t-norm. 
A fuzzy relation with the reflexive and symmetric properties is a {\em proximity} relation. If in addition it has the $\T$-transitive property, it is a {\em similarity} relation.\footnote{
Lexical semantic similarity and a similarity relation are two different concepts. The first simply provides a degree of similarity between words, but it is not a similarity relation  in the sense defined above, with the reflexive, symmetric and $\T$-transitive properties. On the practical side, we use the WordNet similarity measures to obtain the approximation degree that we use when automatically constructing the proximity equations.
}

A weak unification of terms builds upon the notion of \emph{weak unifier} of level $\lambda$ for 
two expressions $\cE_1$ and $\cE_2$ with respect to $\cR$ (or $\lambda$-unifier): 
a substitution $\theta$ such that $\cR(\cE_1\theta,\cE_2\theta) \geq \lambda$, which is the {\em unification degree} of $\cE_1$ and $\cE_2$ with respect to $\theta$ and $\cR$.
There are several weak unification algorithms \cite{JS18b} based on this notion and on the \emph{proximity-based unification relation} $\Rightarrow$, which defines a transition system (based on \cite{MM82}).
This relation, applied to a set of unification problems $\{\cE_i\approx \cE'_i|1\leq i\leq n\}$ can yield either a successful or a failed sequence of transition steps.
In the first case, both a successful substitution and a unification degree are obtained (detailed in, e.g., \cite{JS18b}).
The notion of weak most general unifier (wmgu) $\theta$ between two expressions, denoted by $\wmgu(\cE_1, \cE_2)$, is defined as a $\lambda$-unifier of $\cE_1$ and $\cE_2$ such that there is no other $\lambda$-unifier which is more general than $\theta$.
Unlike in the classical case, the wmgu is not unique. However, our weak unification algorithm computes a representative wmgu with approximation degree greater than or equal to any other wmgu.

Given a fuzzy logic program $\Pi$ with rules $\tuple{(A \leftarrow Q);\delta}$, where $A$ is an atomic formula, $Q$ is either empty or a conjunction of $~n \geq 0~$ atomic formulas $B_i$, and $\delta$ is the degree of the rule, an operational semantics can be defined as a transition system with a transition relation $\sld{}$, which, in particular, includes the  (transition) rule:
%{\mysf Rule 1:} 
\begin{center}
	${\tuple{(\leftarrow \!\!A'\!\wedge\!Q'), \theta, \alpha}}
	\sld{}
	{\tuple{(\leftarrow \!\!Q\!\wedge\!Q')\sigma, \theta\sigma, \delta \T\beta \T\alpha}}$
\end{center}
if $\tuple{(A \leftarrow Q);\delta}\in {\Pi}$,
$\sigma = \wmgu(A, A')\neq fail$, 
$\cR(A\sigma,A'\sigma)=\beta\geq\lambda$, 
and $(\delta \T\beta \T\alpha)\geq \lambda$.

A fuzzy logic program $\Pi$ is translated into a logic program by: linearizing heads, making the weak unification explicit, and explicitly computing the approximation degree. 
Essentially, given a graded rule   
$\tuple{p(\ol{t_n}) \revto Q; \delta}$, 
for each $\cR(p,q)=\alpha\in\Pi$ with $\alpha\geq\lambda$, generate the clause: 
$$
q(\ol{x_n}) 
\revto (\delta\T\alpha) \wedge x_1\approx t_1 \wedge \dots \wedge x_n\approx t_n \wedge Q
$$
\noindent where $\approx$ is the weak unification operator, $t_i$ are terms, $x_i$ are variables, and $\delta\T\alpha$ abbreviates the goal $\delta\T\alpha\geq\lambda$.

We assume 3-arity predicates for lexical similarity measures with the pattern $(w_1\!:\!t_1\!:\!s_1,  w_2\!:\!t_2\!:\!s_2, d)$, where $w_i\!:\!t_i\!:\!s_i$ are \emph{word terms} and $d \in (0,1]$ is a normalized semantic similarity degree. 
Any element $e$ in the semantics of a lexical similarity measure $m$ can be used to generate a proximity equation $\cR(w_1, w_2)=d$ defining $\cR$.
Depending on the fuzzy relation we decide to work with, \BPL\  generates several types of closure starting from the proximity equations defining $\cR$. Specifically,  since a similarity relation requires all of the three properties (in particular, transitivity), its intension is its reflexive, symmetric, $\T$-transitive closure. 
This allows for both manual and automatic generation of proximity equations relating similar words, including words that are not directly related by $m$ (cf. Section \ref{sec-implementation}). 

%%%%%%%%%%%%%%%%%%%%%%%%%%%%%%%%%%%%%%%
Section~\ref{sec-implementation} will show how to integrate WordNet and the aforementioned lexical semantic similarity measures into the state-of-the-art fuzzy logic programming system \BPL.

%%%%%%%%%%%%%%%%%%%%%%%%%%%%%%%%%%%%%%%%%%%%%%%%%%%%%%%%%%%%%%%%
\section{Integrating WordNet into Bousi$\sim$Prolog}\label{sec-implementation}
 
\BPL\footnote{\url{https://dectau.uclm.es/bousi-prolog}} comprises three subsystems with a total of nine modules. 
The \mytt{wn-connect} subsystem provides the basis for the connection between WordNet and the \BPL\ system.\footnote{
\cite{JS19} gives a more detailed description of this subsystem from the user's point of view. Also, \url{https://dectau.uclm.es/bousi-prolog/2018/08/27/applications/} supplies the source files with the code and detailed comments of its implementation.
}
\mytt{wn-connect} is a software application in itself with ten \PL\ modules,
which implements predicates for managing synsets, hypernyms and hyponyms, giving support to the \mytt{wn\_sim\_measures} and \mytt{wn\_ic\_measures} modules which, in addition, implement the standard similarity measures  defined in Section~\ref{sec-WordNet-Similarity}. We now offer a summary of the base modules:
\begin{itemize}
	\item The \mytt{wn\_synsets} module implements predicates to retrieve information about words and synsets stored in WordNet. It uses the \mytt{wn} module implemented by Jan Wielemaker\footnote{\url{https://github.com/JanWielemaker/wordnet}.} which exploits \SWIPL\ demand-loading and Quick Load Files (QLF) for `just-in-time' fast loading.
	
	\item The \mytt{wn\_hypernyms} module implements predicates to retrieve information about hypernyms of a concept (synset). It uses the modules \mytt{wn\_synsets} and \mytt{wn\_utilities}. Notably, the predicate \mytt{wn\_hypernyms/2} returns a list \mytt{List\_SynSet\_HyperNym} of hypernyms (as synset identifiers) for a word term \mytt{Hyponym}. 
	
	\item The \mytt{wn\_hyponyms} module implements predicates to retrieve information about hyponyms of a concept (synset). Remarkably, the predicate \mytt{wn\_gen\_all\_hyponyms\_of/2} generates all the hyponyms of a concept (\mytt{Synset\_ID}), and is especially useful for computing the information content of a concept.
	
\end{itemize}

%%%%%%%%%%%%%%%%%%%%%%%%%%%%%%%%%%%%%%%
\subsection{Implementing Similarity Measures} \label{sec-ImpSimMeasures}

A first step for a more ambitious goal is to automatically extract semantic similarity information from WordNet IS-A hierarchies, and other attributes as the frequency of use as explained before in Section~\ref{sec-WordNet-Similarity}. 
Here, we describe in broad strokes the implementation of similarity measures based on edge-counting (module \mytt{wn\_sim\_measures}) and some insights about those based on information content (module \mytt{wn\_ic\_measures}).

To a greater or lesser extent, all edge-counting similarity measures are based on the computation of the LCS of two words (more accurately,  concepts). The predicate \mytt{wn\_sim\_measures: lcs/6} returns the LCS  of two words \mytt{Word1} and \mytt{Word2}, and also measures depth in their respective HyperTrees. 
Roughly speaking, it computes the HyperTrees of \mytt{Word1} and \mytt{Word2} and compares them from their roots, returning the  \mytt{synset\_ID} previous to the first mismatch (which is the LCS of both concepts). 
Additionally, the predicate \mytt{lcs/6} returns the depths for LCS, \mytt{Word1} and \mytt{Word2} for reasons of efficiency: we want to go through the hypernym lists only once, so these quantities are calculated when computing the LCS.

The computation of the hypernyms of a concept  is carried out by the predicate \mytt{wn\_hypernyms: hypernym\_chain/2}, which computes a list (\mytt{SynSet\_HyperNyms}) of \mytt{synset\_IDs}  designating the hypernyms of a concept (\mytt{SynSet\_Hyponym}). It thus computes a HyperTree that will be used in the former comparison to compute the LCS. 

Once the depths of the LCS and the words to be compared are known, it is easy to compute the relationship degree between them by following the guidelines given in Section~\ref{sec-WordNet-Similarity}. For example, the WUP measure is implemented by the predicate \mytt{wn\_wup/3}, which takes two concepts (expressed as word terms of the form \mytt{Word:SS\_type:Sense\_num}) and returns the degree of similarity between them. It relies on the private predicate \mytt{wup/3} that calls \mytt{lcs/6} to generate and inspect a pair of HyperTrees associated to \mytt{Word1} and \mytt{Word2}, and obtains the similarity degree between both words (according to that pair of HyperTrees). Because a concept can have more than one HyperTree,  several pairs of HyperTrees
are possibly considered, and a list of similarity degrees is obtained 
for each of these pairs of HyperTrees.
Finally, the maximum degree in the list is selected as a result.

Regarding similarity measures based on the information content, the key idea lies in the implementation of the notion of frequency of use. 
The operator \mytt{wn\_s/6} stores information on how common a word is.  The  \mytt{tag\_number} indicates the number of times the word was found in a text corpus: the higher the number, the more
common the word is. This parameter can therefore be employed to obtain the use of a word
and, summing the \mytt{tag\_number} of all words in a synset, the specific 
use associated to a whole synset (i.e., to a concept) can be obtained. 
Then, the frequency of use of a concept is obtained by adding the ``synset tag num'' of all concepts subsumed by  that concept. 

As explained in Section~\ref{sec-WordNet-Similarity}, the information content of a concept is a function of the ratio between the frequency of use of that concept and the frequency of use of the root concept of the hierarchy. Finally, the information-content-based measures are computed as shown in Table~\ref{tab-measures} for specific predicates. 
Note that we have taken the option of smoothing the frequencies of use with a value of 0, which we substitute for a very small number close to 0.
So, some relationship values do not exactly
match those that would be obtained when using tools like \mytt{wordnet::similarity} \cite{PPM04}.

Finally, \ref{sec-eval-measures} includes a performance comparison between 
the similarity measures implemented for \BPL\ and other systems.

%\todo{FSP: Aquí no sé si rige "among" en lugar de "between".
%
%PPP: Creo que es "between" porque en realidad comparamos dos clases de entidades, las que implementa BPL y las que implementan otros.
%
%FSP: Ok.}

%%%%%%%%%%%%%%%%%%%%%%%%%%%%%%%%%%%%%%%
\subsection{Directives for Generating Proximity Equations}
\label{sec-directives}

\BPL\  can load both ontologies (consisting of proximity equations) and fuzzy logic programs (with fuzzy logic rules and possibly proximity equations). Thus, it would be of interest to use the similarity measures implemented in the last section to automatically construct such ontologies.

In order to define the semantic similarity between selected concepts,
we provide a \BPL\ directive for automatically generating the proximity equations which define an ontology:
\begin{itemize}
    \item 
    \mytt{:-wn\_gen\_prox\_equations(+Msr, +LL\_of\_Pats)} \\
    where \mytt{Msr} is the similarity measure which can be any of: \mytt{path}, \mytt{wup}, \mytt{lch}, \mytt{res}, \mytt{jcn} and  \mytt{lin}.   The second argument \mytt{LL\_of\_Pats} is a list for which each element is another list containing the patterns that must be related by proximity equations.
    The pattern can be either a word or the structure \mytt{Word:Type:Sense}, where \mytt{Word} is the word, \mytt{Type} is its type (either \mytt{n} for noun or \mytt{v} for verb), and \mytt{Sense} is the sense number in its synset. 
    Note that, because similarity measures only relate nouns with nouns, and verbs with verbs, the words of a set must be of the same part of speech. If the pattern is simply a word, then a sense number of 1 is assumed, and its type is made to match all other words in the same list.
\end{itemize}

%%\begin{example}
An example of this directive is:
% ex02.bpl
{\myfontcodesize
\begin{verbatim}
:-wn_gen_prox_equations(wup,[[man,human,person],[grain:n:8,wheat:n:2]]).
\end{verbatim}
}

In this case, as only words are provided in the first list, the sense number is 1, and their types are equal by pairs (nouns for these words). 
The second list explicitly specifies the pattern of each word to be related. 
Then, excluding, for reasons of simplicity, reflexive and symmetric entries, the following proximity equations are generated for a 
lambda cut of 0:
{\myfontcodesize
\begin{verbatim}
        sim(man,    human,  1, 0.56).
        sim(man,    person, 1, 0.8888888888888888).
        sim(person, man,    1, 0.8888888888888888).
        sim(human,  person, 1, 0.6086956521739131).
        sim(grain,  wheat,  0, 0.2608695652173913).
\end{verbatim}
}
Note that there are two blocks,
 numbered with 1 for the first four equations, and with 0 for the last one.
Clearly, words in the first list are not made to be related to those in the second list, and therefore they must occur in different blocks. 
In addition, proximity equations are generated only for the words stored in WordNet.
%%\end{example}

Another form of this directive automatically builds an ontology in terms of the tokens in the BPL program by including \mytt{auto} in its second argument. Only the symbols that occur in a program are related, because it would not be practical to relate the symbols of the program with all those that occur in WordNet.

%%%%%%%%%%%%%%%%%%%%%%
\subsection{Implementing the Generation of  Proximity Equations} \label{sec-GenProxEqus}

\BPL\ processes a file (either a program or an ontology) with the load command \mytt{ld \textit{file}} of the \mysf{BPL} Shell module (named \mytt{bplShell}), where its argument is the name of the file to load (with default extension \mytt{bpl}).
Upon execution of this command, a source file (\mytt{\textit{file}.bpl}) is parsed, compiled to Prolog (\mytt{\textit{file}.tpl}), and consulted.

When parsing a directive \mytt{:-wn\_gen\_prox\_equations}, it is first checked for validity, and then replaced in the target \PL\ file with the proximity equations corresponding to the pairs formed with the symbols derived from its arguments.
As explained, there are two cases for this directive, and they are handled in a different way:

\begin{itemize}

\item 
Explicit indication of words to be related.\\
Here, the proximity equations can be directly generated from each list of words, kept in the memory (as asserted \PL\ facts) and outputted to the translated program in the \mytt{.tpl} file at a later stage.
The procedure is as would be expected: for each pair of different words \mytt{W1} and \mytt{W2} in a list, generate the proximity equation \mytt{sim(W1,W2,D)}, where \mytt{D} is the approximation degree for the normalized measure given as the first parameter of the directive.
Normalization is required because measures are generally not on the interval $(0,1]$ which is the range for proximity equations.

\item 
Automatic generation of proximity equations.\\
This case is different from the former because, when processing the directive, the rules in the program have not yet been parsed, so their tokens are not available. 
It is therefore processed after parsing the remaining program, by performing a syntactic analysis in order to extract the sets of constant, functor and predicate identifiers and adding the resulting proximity equations for each separate set of tokens (with the same shape as in the other case) to the memory.

\end{itemize}

The directives that generate proximity equations are based on the private predicate \mytt{gen\_prox\_} \mytt{equation}.
It generates a proximity equation \mytt{sim(Word1, Word2, NormalizedDegree}) in terms of a given measure (\mytt{Measure}) and a pair of words, which can be completely specified with either a pattern or only with its syntactic form as plain words.
In this last case, their first sense number is selected and the same word type is enforced. 

%%%%%%%%%%%%%%%%%%%%%%%%%%%%
\subsection{Accessing WordNet} \label{sec-WN-access}

The \mytt{wn\_connect} subsystem must be made visible before using the built-ins (public predicates) defined in its modules. In \BPL, WordNet and a wide repertoire of built-in predicates which are implemented by the \mytt{wn\_connect} modules can be accessed either by the directive \mytt{:-wn\_connect} in a program or interactively with \mytt{ensure\_loaded(wn(wn\_connect))} at the command prompt.

Nearly all the predicates implemented in the \mytt{wn-connect} subsystem are crisp, returning the top approximation degree. 
For instance, the predicate \mytt{wn\_word\_info/1}  merges the information provided by the predicate \mytt{wn\_s/6} (which stores information about a synset) and \mytt{wn\_g/2} (which contains an explanation/definition of the concept represented by the synset and example sentences). Figure~\ref{fig_word_info_cat} shows the first answer to the query \mytt{wn\_word\_info(cat)}. This is telling us that the first sense (Sense number = 1) of the word form ``cat'' in the part of speech of nouns (Synset type = n —i.e., a noun—) refers to the concept: ``feline mammal usually having thick soft fur and no ability to roar etc.''. There are six more answers for noun-related senses and two more for verb-related senses.

\begin{figure}[t]%[h]
\begin{center}
\begin{verbatim}
?- wn_word_info(cat).
INFORMATION ABOUT THE WORD 'cat' : 
 Synset_id = 102121620
 Word Order num. = 1
 Synset type (n:NOUN, v:VERB, a:ADJ., s:ADJ. SAT., r:ADV.) = n
 Sense number = 1
 Tag_count = 18
 -----------
 Gloss: 
 feline mammal usually having thick soft fur and no ability to roar: domestic 
 cats; wildcats
 -----------
true 
\end{verbatim}
\caption{A query for obtaining all relevant information about a word (e.g. the word ``cat'')}
\label{fig_word_info_cat} 
\end{center}
\end{figure}

\begin{figure}[t]%[h]
\begin{center}
\includegraphics[width=100mm,height=85mm]{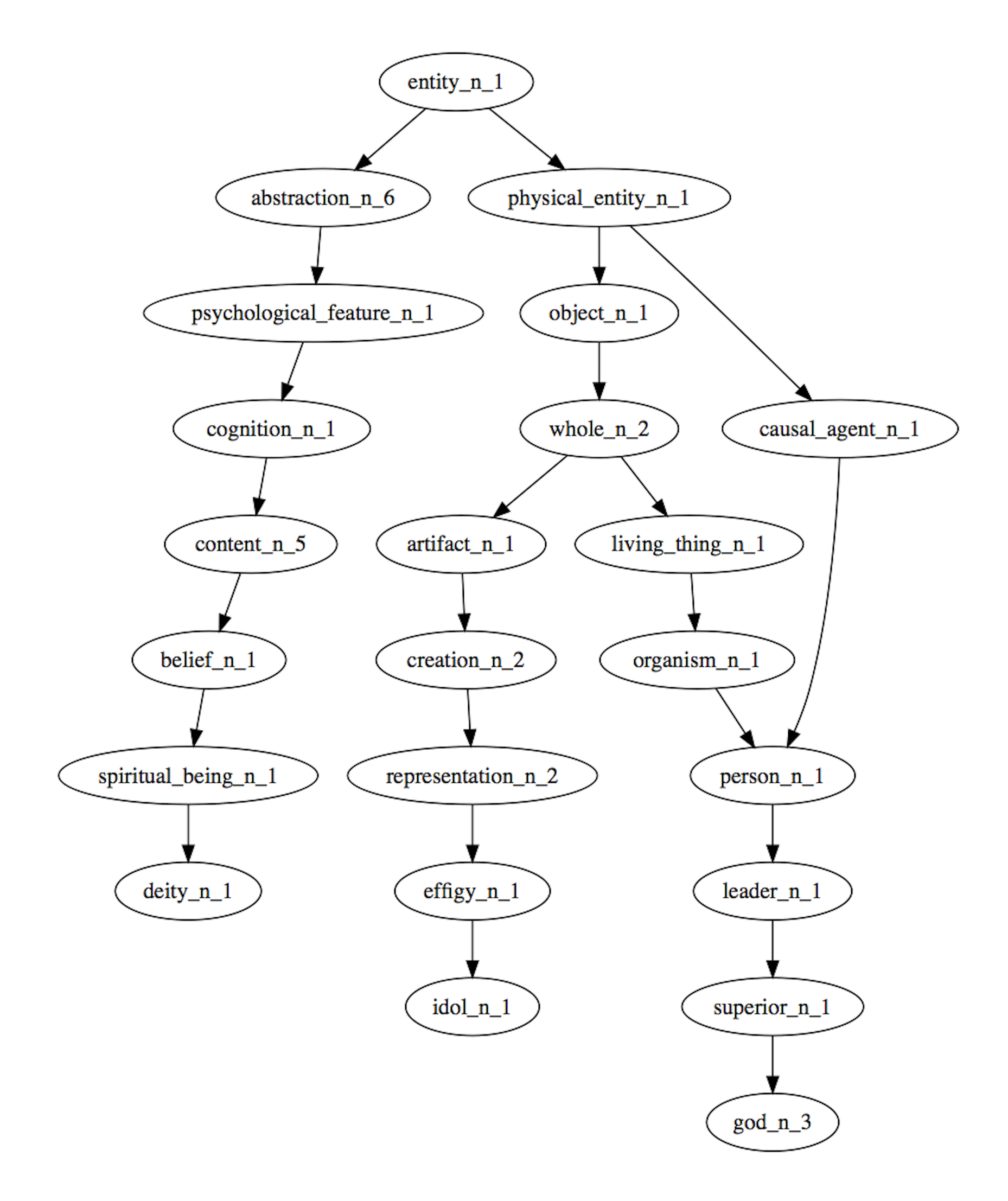}
  \caption{Hypernyms of the word \mytt{god} (all senses)}
  \label{fig_out_god}
\end{center}
\end{figure}

%%%%%%%%
However, the binary similarity predicates (\mytt{wn\_path/2}, \mytt{wn\_wup/2}, \mytt{wn\_lch/2}, etc.) are fuzzy predicates that return the similarity degree of two concepts. 
We also maintain ternary predicates available to programmers, since they provide direct access to the approximation degree \mytt{D}, which can be very useful for its explicit handling. 
Thanks to the repertoire of built-in predicates implemented in the \mytt{wn-connect} subsystem, the user of the \mysf{BPL} system can extract information from WordNet, deepening into the structure of the relationships between its linguistic terms. This becomes especially evident for the predicate \mytt{wn\_display\_graph\_hypernyms/1}. Figure ~\ref{fig_out_god} shows its outcome for the hypernym hierarchy of all senses of the word \mytt{god}.\footnote{
In Figure~\ref{fig_out_god}, each node draws the representative word of the respective synset (i.e., those with   {\tt W\_num} equal to one). This figure also illustrates how a concept can be linked to a hypernym concept through different paths in (a subset of) the WordNet IS-A hierarchy.
}
%

%%%%%%%%%%%%%%%
Moreover, with these built-in predicates, a certain form of linguistic reasoning is possible. For example, 
the predicate \mytt{wn\_lcs/2}, which computes the LCS of a set of concepts, can help to obtain the most specific generalization of a set of concepts and to contribute to knowledge discovery. 
%%%%
%The idea is that although you only have direct information about, e.g., lion, leopard, cougar and cat, in a database, you can discover that this information is also pertinent for feline, using the predicate \mytt{wn\_lcs/2}. 
Although in a database there only exists direct information about, e.g., lion, leopard, cougar and cat, it is possible to discover that this information is also pertinent for feline by using the predicate \mytt{wn\_lcs/2}. 
In Figure~\ref{fig_wn_lcs}, 
the concept referred to by the \mytt{synset\_ID 102120997} (grouping \mytt{[feline:n:1, felid:n:1]}) is the most specific concept of (the synsets of) \mytt{lion}, \mytt{leopard}, \mytt{cougar} and \mytt{cat}, that they share as a common ancestor in the IS-A hierarchy of WordNet.
\begin{figure}[t]%[h]
\begin{center}
\begin{verbatim}
?- wn_lcs([lion, leopard, cougar, cat], LCS_SS_ID), 
   wn_synset_components(LCS_SS_ID, Words_LCS_SS_ID).
LCS_SS_ID = 102120997,
Words_LCS_SS_ID = [feline:n:1, felid:n:1].
\end{verbatim}
\caption{A query for obtaining the LCS or most specific generalization of a set of concepts.}
\label{fig_wn_lcs}
\end{center} 
\end{figure}
%%%%
Furthermore, the predicate \mytt{wn\_gen\_hyponyms\_upto\_level/3}, which generates all the hyponyms of a concept (\mytt{Synset\_ID}) up to a certain depth level (\mytt{Level}), can also be used to generate an ontology of closely related terms to the given concept that can be used to implement flexible queries and text mining tasks.
In particular, \ref{sec-Text-Class} illustrates an application of this work to text classification, also including some performance measures.

%%%%%%%%%%%%%%%%%%%%%%%%%%%%
\section{Experimental Assessment} \label{sec-EXP-assessment}

In the following two subsections we perform experiments to find the performance of the implemented similarity measures and the cost of integrating WordNet into \BPL.
Instructions, programs and data to reproduce the experiments in these appendices have been made available at \url{https://dectau.uclm.es/bousi-prolog/wp-content/uploads/sites/3/2020/07/Published.zip}.

%%%%%%%%%%%%%%%%%%%%%%%%%%%%%%%%%%%%%%%%%%%%%%%%%%%%%%%%%%%%%%
%%%%%%%%%%%%%%%%%%%%%%%%%%%%%%%%%%%%%%%%%%%%%%%%%%%%%%%%%%%%%%
\subsection{Evaluation of the Measures and Comparison with other Systems} \label{sec-eval-measures}

In this section we evaluate the computational cost of the measures implemented in Subsection~\ref{sec-ImpSimMeasures}, comparing the results with other systems.

Specifically,  we are using an implementation of WordNet::Similarity for Java (WS4J) developed by Hideki Shima.\footnote{%
WS4J is available at \url{https://github.com/Sciss/ws4j} and also has a web interface WS4J Demo at
\url{http://ws4jdemo.appspot.com}.
} We use WS4J because it provides some time information that allows the cost of these measures to be appreciated. We are also using our own implementation of the WordNet-based similarity measures, but executed both by \SWIPL\ and \BPL. This allows us not only to compare the performance of our measures integrated into \BPL\ with WS4J, but also the overhead introduced by our implementation of \BPL\ w.r.t. \SWIPL.

In the first experiment, we selected twelve words with the highest number of senses. Then we pair them obtaining six pairs of words. Afterwards, for each word in that pair we generate the Cartesian product of all their senses (\mytt{Word1:n:Sense1}, \mytt{Word2:n:Sense2}). Finally, we compute the similarity degree of these pairs, thus mimicking how WS4J operates,\footnote{For WS4J, the parameter MFS (Most Frequent Sense) is set to false.} measuring the overall time cost of the computation.

Table~\ref{tab-WS4J-BPL-SWIPL} shows the costs involved in the computation of the similarity degree of these six pairs of words for the three systems and the three edge-based measures.
For each measure, `Time' shows the elapsed time in milliseconds, `Lat.' the latency in milliseconds/pair, and `Thr.' the throughput in pairs/second.
While BPL is at a small disadvantage with respect to SWIPL, WS4J is roughly four times as fast.

\begin{table}[htdp]
	\caption{Comparing \BPL, \SWIPL\ and WS4J on six pairs of words with the highest number of senses}
	\label{tab-WS4J-BPL-SWIPL}
	%\begin{minipage}{\textwidth}
		\begin{tabular}{|l@{}r|@{}r|@{}r|@{}r|@{}r|@{}r|@{}r|@{}r|@{}r|}
			\cline{1-10}
			\multicolumn{10}{@{}c}{\vspace{-1em}}\\
			\cline{1-10}
             & \multicolumn{9}{|@{}c|}{Measure}\\
			\cline{2-10}
			System & \multicolumn{3}{|c|}{PATH} & \multicolumn{3}{@{}c|}{WUP} & \multicolumn{3}{@{}c|}{LCH} \\ 
			\cline{2-10}
			&  \multicolumn{1}{|c|}{Time} & \multicolumn{1}{c|}{Lat.} & \multicolumn{1}{c|}{Thr.} & Time & Lat. & \multicolumn{1}{c|}{Thr.} & Time & Lat. & \multicolumn{1}{c|}{Thr.} \\
			\cline{1-10}
			BPL         & \multicolumn{1}{|r|}{966} & 0.02& 44,572 & 959 & 0.02 & 45,400 & 959 & 0.02& 44,943	\\  
			SWIPL       & \multicolumn{1}{|r|}{895}	& 0.02 & 48,004	& 920 & 0.02 &	46,540 & 950 & 0.02 & 45,117 \\   
			WS4J & \multicolumn{1}{|r|}{211}& 0.01 & 204,410 & 430& 0.01 & 101,024 & 212 & 0.01 & 205,464 \\
			\cline{1-10}
			\multicolumn{10}{@{}c}{\vspace{-1em}}\\
			\cline{1-10}
		\end{tabular}
	%\end{minipage}
\end{table}%

In a second experiment, we randomly generate pairs of noun and verb patterns (\mytt{Word1:Type:} {\tt Sense1}, \mytt{Word2:Type:Sense2}) so \mytt{Type} is either \mytt{n} or \mytt{v}. Then, we generate the calls to a similarity measure, and finally, we measure the performance of \BPL\ w.r.t. \SWIPL.\footnote{\url{https://code.google.com/archive/p/ws4j/wikis/DraftNextVersion.wiki} describes a similar experiment for WS4J, but we were unable to replicate it because the results of this kind of experiments depends strongly on the list of word pairs.} Table~\ref{tab-BPL_vs_SWIPL} shows the results of this experiment. The numbers are the average after tree executions.

\begin{table}[htdp]
\caption{Comparing \BPL\ and \SWIPL\ on randomly generated pairs of patterns}
\label{tab-BPL_vs_SWIPL}
\begin{minipage}{\textwidth}
\begin{tabular}{|lr|r|r|r|r|r|}
\cline{1-7}
\multicolumn{7}{c}{\vspace{-1em}}\\
\cline{1-7}
             & \multicolumn{6}{|c|}{Measure} \\ 
\cline{2-7}
System     & \multicolumn{3}{|c|}{Edge-based} & \multicolumn{3}{c|}{Information Content-based}\\
\cline{2-7}
                &  \multicolumn{1}{|c|}{PATH}     & WUP     & LCH      & RES      & JCN      & LIN\\
                & \multicolumn{3}{|c|}{(milliseconds/10,000 pairs)} & \multicolumn{3}{c|}{(milliseconds/250 pairs)}\\
\cline{1-7}
BPL         &  \multicolumn{1}{|r|}{1,438}     & 1,403     & 1,597      & 34,680      & 36,221      & 34,982\\     
SWIPL    &  \multicolumn{1}{|r|}{242}     & 242     & 275      & 33,521      & 35,088      & 35,073\\
\cline{1-7}
\multicolumn{7}{c}{\vspace{-1em}}\\
\cline{1-7}
\end{tabular}
\end{minipage}
\end{table}%

The analysis of the data in Table~\ref{tab-BPL_vs_SWIPL} reveals that for the BPL system the average Latency of edge-based measures is 
%0.14798 
0.14 ms/pair and the average Latency of IC-based measures is 
%141.17778 
141.17
ms/pair, while for \SWIPL\ they are 
%0,02532 
0.03 ms/pair and 
%138,24311
138.24 ms/pair respectively. These results lead to an average ratio between both systems of 5.84 for edge-based measures and only 1.02 for IC-based measures, showing an acceptable overhead of \BPL\ relative to \SWIPL\ for these tasks.
In the first case, the overhead is more noticeable when traversing 10K word pairs than only 250 because tail recursion optimization is lost in the \BPL\ to Prolog translation.
Thus, optimizing this translation will be the subject of future work.

%%%%%%%%%%%%%%%%%%%%%%%%%%%%%%%%%%%%%%%

%%%%%%%%%%%%%%%%%%%%%%%%%%%%%%%%%%%%%%%%%%%%%%%%%%%%%%%%%%%%%%
%%%%%%%%%%%%%%%%%%%%%%%%%%%%%%%%%%%%%%%%%%%%%%%%%%%%%%%%%%%%%%
\subsection{Applications to Text Classification and some Performance Results} \label{sec-Text-Class}

\BPL\ is well suited to making the query-answering process more flexible, due to its weak unification 
algorithm. In \cite{RJ14JIFS} we discussed several practical applications where it can be useful, such as
flexible deductive databases, knowledge-based systems, information retrieval, and approximate reasoning.
\BPL\ has been used in a number of real applications such as:
text classification (or cataloging) \cite{RJFG13JLRE}, knowledge discovery \cite{RJ15JIFS},  
linguistic feedback in computer games \cite{RT16}, and decision making \cite{CU19,CU20}.

In this section we briefly summarize our latest research in text classification. The goal of any text classification process is to assign one or more predetermined categories to classify each of the texts. We are proposing a declarative approach consisting of classifying texts according to a set of predefined categories by using semantic relations and the ability of \BPL\ to weakly unify. The proposed method consists of the following steps: 
\begin{enumerate}
 \item \textbf{Knowledge Base Building}: 
 The categories are (semantically) defined by extracting a set of proximity equations from standard thesauri and ontologies (WordNet in our case). The set of proximity equations form the significative subset of the thesaurus or ontology that we will use in the classification process and, by abuse of language, we name it the ``ontology'' file.
 
 \item \textbf{Flexible Search and Computing Occurrence Degrees}: 
 For each document content, the words close to a category are searched in order to classify them, and their degrees of occurrence are obtained.  The \emph{occurrence degree} of a word is an aggregation of the number of occurrences
of the word (in a document)  and its approximation  degree with regard to the category analyzed.

 \item \textbf{Computing Document Compatibility Degrees}: 
 The compatibility degrees of the documents with regard to a category are computed using a specific compatibility measure.   A \emph{compatibility measure} is an operation which uses the occurrence degrees of the words close to a category to calculate a document compatibility degree, that is, an index of how compatible the document is with regard to the category analysed.
 
 \item \textbf{Classification Process}: 
 Finally, each document is classified as pertaining to the category or categories that return a higher compatibility degree. We assign to a document all the categories that have a compatibility index between the maximum compatibility, {\tt Max}, obtained for that document and a minimum {\tt Min=0.9*Max}.
\end{enumerate}

It is noteworthy that our approach to text classification does not need a pre-classified set of training documents. The proposed method only requires the category names as user input. Hence, our method is not based on category occurrence frequency, but depends greatly on the definition of that category and how the text fits that definition.

Thanks to the integration with WordNet, we can generate the ontology files without human intervention, starting from the set of predefined categories.  Ontology files are computed either by: i) generating several level of hyponyms of a category and obtaining the similarity degree between them and the category by using a similarity measure (PATH, WUP, etc.); or ii) taking the gloss of a category (which can be seen as the definition of the category), extracting a list of words using natural language processing techniques, and then obtaining the degree of relation between them and the category by using a similarity measure.

Once the ontology file is generated and the categories from which we start are defined (in terms of their semantic relationship with other words), we can then apply the remaining steps of our classification algorithm.

An application implementing the method described above can be found at the URL \url{https://dectau.uclm.es/bousi-prolog/applications/}, and a preliminary paper on this subject is \cite{AJRS20}. The results shown in that paper are encouraging in terms of \emph{Precision}, \emph{Recall} and \emph{F-measure}.\footnote{
\emph{Precision}: percentage of total positive classifications w.r.t. the total of classifications performed by the classifier method. In this case, 'positive classification' means a classification where the classifier and the expert judgment coincide.
\emph{Recall}:   percentage of total positive classifications w.r.t. the total of classifications 
performed by the expert classifier.
\emph{F-measure}: the harmonic mean between precision and recall.
} 
For instance, for the dataset ``News Wires-2 (Reuters-10)'', 
which is a set of short texts (news limited up to 160 characters long) selected from Reuters-21578,\footnote{
\url{http://www.daviddlewis.com/resources/testcollections/reuters21578/}
}
we obtain an average \emph{Precision}, \emph{Recall} and \emph{F-measure} of 73.07\%, 55.59\% and 62.99\% respectively. Our immediate goal is to improve \emph{Recall} and to contribute to provide explainable results.

In order to show the feasibility of integrating WordNet into \BPL, we undertook an experimental assessment of the cost of generating several ontologies and classifying some datasets. The results are shown in Tables \ref{tab-ontologies2} and \ref{tab-classifying2},
with CPU runtime in seconds, the number of inferences performed during the computation, and information on memory consumption in megabytes. 

\begin{table}[ht]
\caption{Performance of automatic generation of ontologies based on WordNet hyponyms}\label{tab-ontologies2}
\begin{minipage}{\textwidth}
\begin{tabular}{|l|r|r|r|r|r|}
\cline{1-6}
\multicolumn{6}{c}{\vspace{-1em}}\\
\cline{1-6}
Ontology file &  Equs. &  Runtime (s)  & Inferences   & Global Stack (Mb) &    Local Stack (Mb) \\
\cline{1-6}
\multicolumn{6}{|c|}{Using similarity measures based on counting edges}\\
\cline{1-6}
odp\_hyp 			&263 	&0.038	&287,140 		&0.823461 	&0.512711 \\
enviweb\_hyp  		&352 	&0.486 	&399,114		&0.767675 	&0.723183 \\
reutersshorts\_hyp	&278		&0.039	&286,338		&0.348183	&0.554588 \\
reuters10\_hyp		&314		&0.044	&328,070		&0.541951	&0.618492 \\
\cline{1-6}
AVERAGE		&302		&0.152	&325,165		&0.620317	&0.602243 \\
\cline{1-6}
\multicolumn{6}{|c|}{Using similarity measures based on information content}\\
\cline{1-6}
odp\_hyp\_ic 			&263 	&44.020 	&190,808,568 	&0.245071 	&0.512710 \\
enviweb\_hyp\_ic		&352 	&28.529 	&124,793,835 	&0.344101 	&0.723183 \\
reutersshorts\_hyp\_ic 	&278		&1.641	&7,382,626 	&0.807327 	&0.554588 \\
reuters10\_hyp\_ic 		&314 	&25.871 	&111,821,332 	&0.323959	&0.618492 \\
\cline{1-6}
AVERAGE			&302		&25.015	&108,701,590	&0.430115	&0,602243 \\
\cline{1-6}
\multicolumn{6}{c}{\vspace{-1em}}\\
\cline{1-6}
\end{tabular}
\end{minipage}
\end{table}

\begin{table}[ht]
%\caption{Performance of classifying datasets from WordNet automatically generated ontologies}\label{tab-classifying2}
\caption{Performance of classifying datasets from WordNet-generated ontologies}\label{tab-classifying2}
\begin{minipage}{\textwidth}
\begin{tabular}{|l|r|r|r|r|}
\cline{1-5}
\multicolumn{5}{c}{\vspace{-1em}}\\
\cline{1-5}
Dataset &  Runtime (s)  & Inferences   & Global Stack (Mb) &    Local Stack (Mb) \\
\cline{1-5}
\parbox[c]{3.5cm}{\vspace*{0.5mm}
	Web Snippets  (ODP): $115$ documents\vspace*{0.5mm}} 	&1.453	&14,510,417	&2.49	&2.31\\[1ex]
%Web Snippets  (ODP): $115$ documents 	&1.453	&14,510,417	&2.49	&2.31\\[2ex]
\cline{1-5}
\parbox[c]{3.5cm}{\vspace*{0.5mm}News Snippets (EnviWeb): $116$ documents\vspace*{0.5mm}} 	&1.668	&16,508,815	&1.72	&1.71\\[0.75ex]
\cline{1-5}
\parbox[c]{3.5cm}{\vspace*{0.5mm}News Wires-1 (Reuters-Short): $267$ documents\vspace*{0.5mm}}		&2.766	&27,308,333	&3.32	&4.39\\[0.75ex]
\cline{1-5}
\parbox[c]{3.5cm}{\vspace*{0.5mm}News Wires-2 (Reuters-10): $8.599$ documents\vspace*{0.5mm}}	&422.729	&3,586,006,321	&247.62	&341.49\\[0.75ex]
\cline{1-5}
\multicolumn{5}{c}{\vspace{-1em}}\\
\cline{1-5}
\end{tabular}
\end{minipage}
\end{table}

In Table~\ref{tab-ontologies2}, each row groups the average data obtained when generating ontologies for a given data set using three different similarity measures. The column ``Equs.'' shows the number of proximity equations generated per ontology. The first part of the table presents data related to similarity measures based on counting edges (PATH, WUP and LCH) while the second part gives those based on information content (RES, JCN, LIN). 

Note that, for the ontologies which use similarity measures based on counting edges, the cost of generating and storing proximity equations (pairs) in a file 302 is 0.152 seconds, on average. This signifies that for this kind of measure the latency is 0.5 ms/equ and the throughput 1,986.8 equs/s. Similarly, for measures based on information content, the latency is 82.9 ms/equ, and the throughput is 12.1 equs/s. 

As can be seen, building ontologies using similarity measures based on information content has a higher cost due to the complexity of this kind of similarity measure (involving the computation of the LCS and the generation of all its hyponyms, to establish its information content, in order to compute the similarity degree of two words).

Table~\ref{tab-classifying2} sets out information about the average cost of classifying four different datasets using the previously generated ontologies.

%%%%%%%%%%%%%%%%%%%%%%%%%%%%%%%%%%%%%%%%%%%%%%%%%%%%%%%%%%%%%%%
%%%%%%%%%%%%%%%%%%%%%%%%%%%%%%%%%%%%%%%%%%%%%%%%%%%%%%%%%%%%%%%
\section{Conclusions} \label{sec-conc}

We have presented techniques to embody the information stored in the lexical database WordNet \cite{Fel98,Fel06,Mil95} into the fuzzy logic programming language \BPL\ \cite{RJ14JIFS,JR17FSS,JS18b}. However, the techniques developed can be used to connect WordNet to any logic programming language that uses an operational semantics based on some variant of WSLD resolution.

The main contributions of this study are the following:
\begin{enumerate}
\item We have implemented, in Prolog, all the usual similarity measures (based on counting edges and on information content) to be found in standard tools such as \mytt{wordnet::simila-} \mytt{rity} \cite{PPM04}.

\item A whole \mysf{BPL} subsystem (\mytt{wn-connect}) has been developed, providing those measures and several built-in predicates to obtain useful information about words and synsets in WordNet. This subsystem can be used independently in a \PL\ interpreter. 

\item 
We have implemented directives to generate proximity equations from a set of words, linking them with an approximation degree. Hence, the significance of this work is to make a fuzzy treatment of concepts via proximity relations possible, and also to endow \BPL\ with linguistic characteristics.

\item Because \BPL\ allows WordNet databases to be accessed easily, it is possible to use interesting relations (antonymy, meronymy, etc.), or to use causal relations, to reason.

\item We have provided the system implementing these techniques as a desktop application (for Windows, Mac and Linux OS's -- \url{dectau.uclm.es/bousi-prolog}), and also an online web interface (\url{dectau.uclm.es/bplweb}).

\item Finally, we have undertaken an experimental assessment: Firstly, measuring the performance of the implemented WordNet-based similarity measures and the cost of generating hyponymy-based ontologies; and, secondly, executing a text classification application implemented using \BPL\ and its connection to WordNet, concluding that \BPL\ has a reasonable performance w.r.t. other systems.
\end{enumerate}

As future work, experiments suggest enhancing the performance of \BPL\ by introducing memorizing techniques, and optimizing their compilation by leveraging tail recursion optimization, and also, implementing relatedness measures based on other techniques such as word embeddings.

\subsection*{Acknowledgements}
{\small
We would like to express our gratitude to the anonymous reviewers and the area editor for their valuable comments that have greatly improved the final version of our paper.
}

%%%%%%%%%%%%%%%%%%%%%%%%%%

%%%%%%%%%%%%%%%%%%%%%%%%%%%%%%%%%%%%%%%%%%%%%%%%%%%%%%%%%%%%%%%
%%%%%%%%%%%%%%%%%%%%%%%%%%%%%%%%%%%%%%%%%%%%%%%%%%%%%%%%%%%%%%%
%%%%%%%%%%%%%%%%%%%%%%%%%%%%%%%%%%%%%%%%%%%%%%%%%%%%%%%%%%%%%%%
% \section*{Appendices}

% \appendix

%%%%%%%%%%%%%%%%%%%%%%%%%%%
% \section{Appendix-1} \label{sec-Appendix-1}

%%%%%%%%%%%%%%%%%%%%%%%%%%%%%%%%%%%%%%%
\bibliographystyle{acmtrans}
%\bibliography{biblioPaper}

%%%%%%%%%%%%%%%%%%%%%%%%%%%%%%%%%%%%%%%

\label{lastpage}
\end{document}